\documentclass[lettersize,journal]{IEEEtran}
\usepackage{amsmath,amsfonts}
\usepackage{algorithmic}
\usepackage{algorithm}
\usepackage{array}
\usepackage[caption=false,font=normalsize,labelfont=sf,textfont=sf]{subfig}
\usepackage{textcomp}
\usepackage{stfloats}
\usepackage{url}
\usepackage{verbatim}
\usepackage{graphicx}
\usepackage{cite}

\usepackage{bm}
\usepackage{amsmath}

\begin{document}

\title{Performance Analysis of 6TiSCH Networks\\Using Discrete Events Simulator}

\author{Guilherme~de~Santi~Peron, Marcos~Eduardo~Pivaro~Monteiro,
João~Luís~Verdegay~de~Barros, Jamil~Farhat and~Glauber~Brante
\thanks{G.S. Peron and J.L.V. Barros are with Graduate Program in Energy Systems (PPGSE), Federal University of Technology - Paraná (UTFPR), Curitiba-PR, Brazil (e-mails: peron@utfpr.edu.br, barros.joao.luis@gmail.com)\\
M.E.P. Monteiro is with Academic Department of Electronics (DAELN), Federal University of Technology - Paraná (UTFPR), Curitiba-PR, Brazil (e-mail: marcose@utfpr.edu.br)\\
J. Farhat and G. Brante are with Graduate Program in Electrical and Computer Engineering (CPGEI), Federal University of Technology - Paraná (UTFPR), Curitiba-PR, Brazil (e-mails: jamilfarhat@utfpr.edu.br, gbrante@utfpr.edu.br)}}

\maketitle

\begin{abstract}
The Internet of Things (IoT) empowers small devices to sense, react, and communicate, with applications ranging from smart ordinary household objects to complex industrial processes. To provide access to an increasing number of IoT devices, particularly in long-distance communication scenarios, a robust low-power wide area network (LPWAN) protocol becomes essential. A widely adopted protocol for this purpose is 6TiSCH, which builds upon the IEEE 802.15.4 standard. It introduces time-slotted channel hopping (TSCH) mode as a new medium access control (MAC) layer operating mode, in conjunction with IEEE 802.15.4g, which also defines both MAC and physical layer (PHY) layers and provides IPv6 connectivity for LPWAN. Notably, 6TiSCH has gained adoption in significant standards such as Wireless Intelligent Ubiquitous Networks (Wi-SUN). This study evaluates the scalability of 6TiSCH, with a focus on key parameters such as queue size, the maximum number of single-hop retries, and the slotframe length. Computational simulations were performed using an open-source simulator and obtained the following results: increasing the transmission queue size, along with adjusting the number of retries and slotframe length, leads to a reduction in the packet error rate (PER). Notably, the impact of the number of retries is particularly pronounced. Furthermore, the effect on latency varies based on the specific combination of these parameters as the network scales. 
\end{abstract}

\begin{IEEEkeywords}
6TiSCH, TSCH, LPWAN, IoT.
\end{IEEEkeywords}

\section{Introduction}
\IEEEPARstart{E}{mpowered}  by sensing, communication, and computing capabilities, the Internet of Things (IoT) stands as a revolutionary technology that seamlessly bridges the physical and virtual worlds. Projections indicate that by $2030$, approximately $80$ billion IoT devices will be interconnected~\cite{Chettri.IoT.2020}. Leveraging advancements in communications, edge computing, and artificial intelligence, IoT has enabled a multitude of applications and services, including smart homes, agriculture, smart cities, and more~\cite{Keersmaeker.2023, Wang.2022}. Within this context, a reliable low-power wide-area network (LPWAN) protocol is crucial for communication over long distances, as seen in scenarios like smart grid solutions~\cite{Venturini.2022}. Then, in this variety of application scenarios, IEEE 802.15.4g stands out as one very promising protocol stack. It utilizes the upper layers defined by IPv6 over the time-slotted channel hopping (TSCH) mode of IEEE 802.15.4 (6TiSCH), which is supported by the Internet Engineering Task Force (IETF). This long-range mesh protocol has been considered in various works~\cite{Yadav.2023, Melo.2023, Kim.2017}.

Considering the extensive coverage of LPWA networks, each node faces the challenge of sharing the communication spectrum with numerous other communication devices. Despite various available strategies to mitigate interference, such as employing orthogonal configurations of center frequency, spreading factor (SF), and similar parameters, a limitation remains regarding the maximum feasible number of nodes within a designated area. This scalability issue of the 6TiSCH stack protocol has been investigated in~\cite{Barros.2022}, revealing that the 6TiSCH network experiences a significant impact in terms of packet error rate (PER) as the number of nodes increases. This impact is due to effects like collisions and full transmit queues of the IoT devices.

\subsection{Related Work}
\label{subsec:RelatedWork}

Numerous studies have investigated network performance involving IPv6 over IEEE 802.15.4 TSCH and 6TiSCH. Specifically, Kim \textit{et al.}~\cite{Kim.2017} examined the constraints associated with deploying a large network using IPv6 over IEEE 802.15.4 TSCH. They proposed an enhanced scheme for configuration management and scheduling. However, their work lacks traffic tests to evaluate the effectiveness of their proposed solution. Furthermore, concerning 6TiSCH, Accettura \textit{et al.}~\cite{Accettura.2015} implemented this protocol in a factory automation use case using the OpenWSN protocol stack. They successfully addressed scalability issues related to multi-hop dense low-power networks. However, it is important to note that their study was confined to the factory automation domain and did not encompass other LPWAN scenarios.

Considering an empirical methodology, Watteyne \textit{et al.}~\cite{Watteyne.2015} documented the connectivity of a TSCH network with $350$ nodes in an office environment, covering $16$ frequencies at the $2.4$~GHz band. Notably, the study revealed the significant impact of communication frequency on the quality of individual links. However, it was restricted to the $2.4$~GHz band, which limits the maximum single-hop range and is unsuitable for LPWAN protocols. Barros \textit{et al.}~\cite{Barros.2022} conducted an analysis of the scalability of 6TiSCH and LoRAWAN for LPWAN. However, this study did not explore crucial parameters, including queue size, maximum single-hop retries, and the length of the slotframe.

\subsection{Novelty and Contribution}
\label{subsec:Novelty}

In this work, differently from~\cite{Barros.2022}, in which a comparison between LoRaWAN and 6TiSCH is performed, we investigate the impact of queue size, maximum single-hop retries, and slotframe length focusing on the scalability of the 6TiSCH stack protocol. To the best of our knowledge, this is the first work that explores these parameters from a scalability perspective, which constitutes the novelty of this work. To obtain such results, the open source 6TiSCH Simulator~\cite{Municio.2019} is employed to analyze the network. The results show that increasing the maximum number of single-hop transmission attempts exerts substantially more influence on system scalability than enlarging the transmission queue size. Additionally, we observe that increasing the transmission queue size can reduce both latency and PER when the slotframe is expanded. 

\section{System Model}
\label{sec2}

\begin{figure}[!htb]
\center{
\includegraphics[width=0.9\columnwidth]{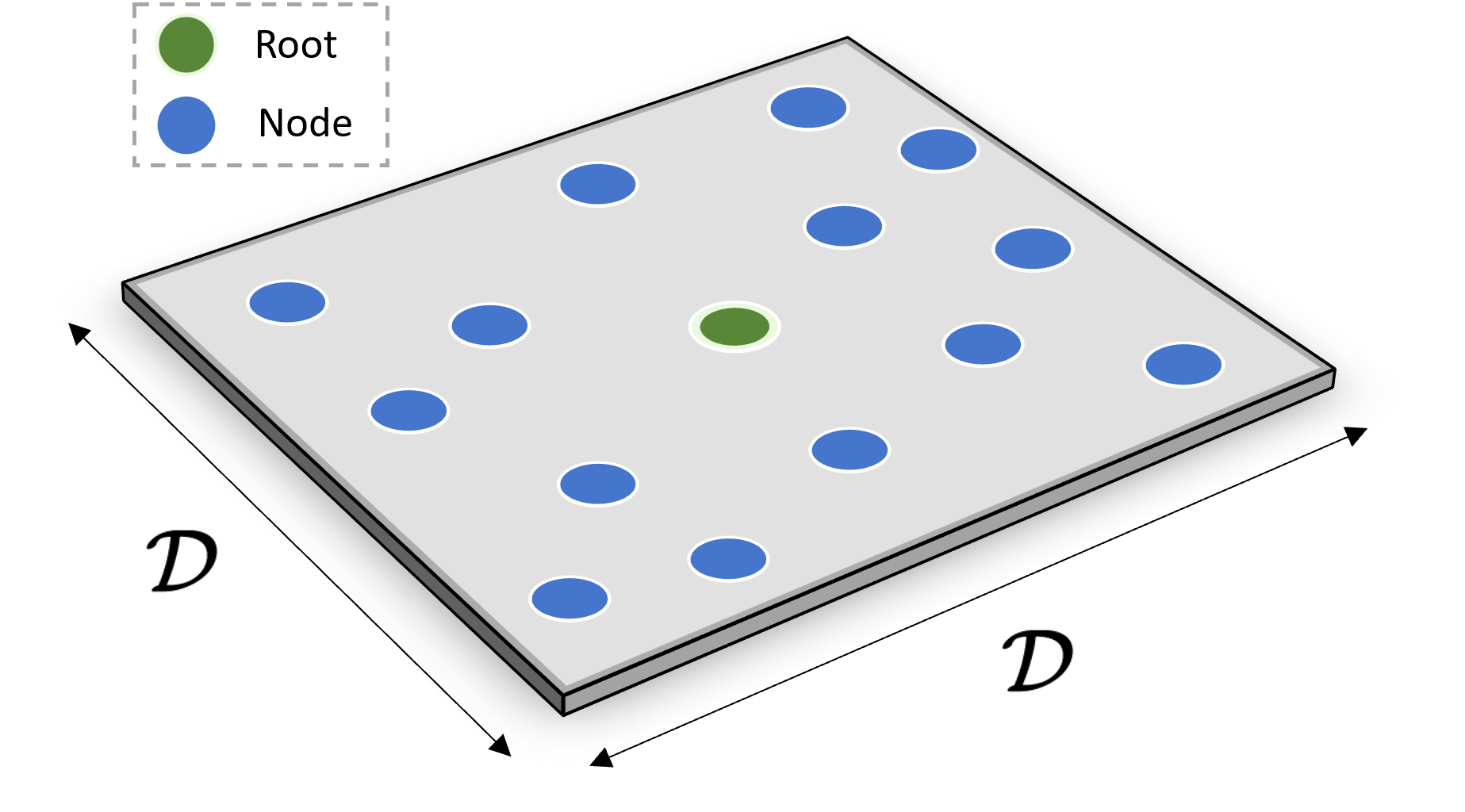}}
\caption{System model with $\mathcal{N}$ nodes uniformly distributed in a square area of $\mathcal{D} \times \mathcal{D}$ $\text{km}^2$ with a root node at the center.}
\label{fig:systemmodel}
\end{figure}

We consider $\mathcal{N}$ nodes uniformly distributed in a square area of $\mathcal{D} \times \mathcal{D}$ $\text{km}^2$ that transmit messages to a root node at the center of the square, as shown in Fig.~\ref{fig:systemmodel}. All nodes operate in the license-free $915$~MHz industrial scientific and medical (ISM) band to enable long-range transmissions and avoid frequency-related fees. The channel spacing is $\mathcal{S} = 200$~kHz, which is a common parameter for 802.15.4g using operating mode $1$ for the US ISM ($902$-$928$~MHz) frequency band~\cite{IEEE.2012}.

Using the Friis model, the received power, $\mathcal{P}_\mathrm{r}$, in dBm is defined as~\cite{Goldsmith.05}
\begin{equation}
\begin{split}
\mathcal{P}_{\mathrm{r}} = 10 \log_{10} \left( \frac{\mathcal{P}_\mathrm{t} \mathcal{G} \lambda^2}{\left( 4\pi d \right)^2 } \right), 
\label{eq:pfriss}
\end{split}
\end{equation}
where $d$ is the link distance in meters, $\mathcal{P}_\mathrm{t} = 14$~dBm is the transmitted power\footnote{$\mathcal{P}_\mathrm{t} = 14$~dBm is the maximum transmit power for many LPWAN microcontrollers~\cite{CC1352.2020}, while also being the maximum output power allowed by the European Telecommunications Standards Institute (ETSI).}, $\lambda$ is the wavelength in meters, and $\mathcal{G} = \frac{\mathcal{G}_\mathrm{t} \mathcal{G}_\mathrm{r}}{L}$ is the gain associated with transmitting antenna ($\mathcal{G}_\mathrm{t}$), receiving antenna ($\mathcal{G}_\mathrm{r}$) and loss factor ($L$), which are all assumed to be unitary in this work.

We employ the Pister-Hack propagation model to account for long-term variations in transmission. In this model, the Received Signal Strength Indicator (RSSI) levels are fine-tuned to match with empirical results~\cite{Elsts.2020}. Specifically, we subtract a uniformly distributed variable ranging from $0$ to $40$~dB from the expression in~\eqref{eq:pfriss}. As a result, for a received power level of $\mathcal{P}_\mathrm{r} = -80$~dBm, the final RSSI spans from $-80$~dBm to $-120$~dBm.

\section{Results}
\label{sec3}

In this section, we evaluate the performance of 6TiSCH under various network parameters. We focus on how the maximum retries, queue size, and slotframe length affect the PER and latency. We consider two slotframe lengths: $\mathcal{T} = 101$~timeslots, with a slot duration of $0.04$~s, which is the default in 6TiSCH~\cite{RFC8180}, resulting in a slotframe of $4.04$~s; and $\mathcal{T} = 606$~timeslots, with a slot duration of $0.04$~s, resulting in a slotframe of $24.24$~s.  Consequently, the slotframes repeat approximately every $4$~s and $24$~s, respectively, for $\mathcal{T} = 101$ and $\mathcal{T} = 606$. Additionally, we assume that $\mathcal{D} = 2$~km, and that $\mathcal{C} = 8$ channels are available, following the same assumption as Barros \textit{et al.}~\cite{Barros.2022}.

Each node's position is uniformly randomly generated in the available area, and there is a warm-up period in which the simulation awaits all the nodes to join the network. Furthermore, we simulate a network based on the 6TiSCH architecture using the 6TiSCH Simulator with its standard parameters, as described by~\cite{Barros.2022}. This simulation allows us to perform specific parameter comparisons. Additionally, the network employs the Routing Protocol for Low Power Lossy Networks (RPL), utilizing the default Objective Function Zero (OF$0$) implementation within the 6TiSCH Simulator to optimize multipoint-to-point routes. The packet transmission rate per node is Poisson distributed with, on average, one packet every $t = 1$ minute. Such update is similar to that used in \cite{Palattella.2016} and corresponds to situations like that seen in WirelessHART networks. 

\begin{figure}[!t]
    \centering
    \includegraphics[width=1.0\columnwidth]{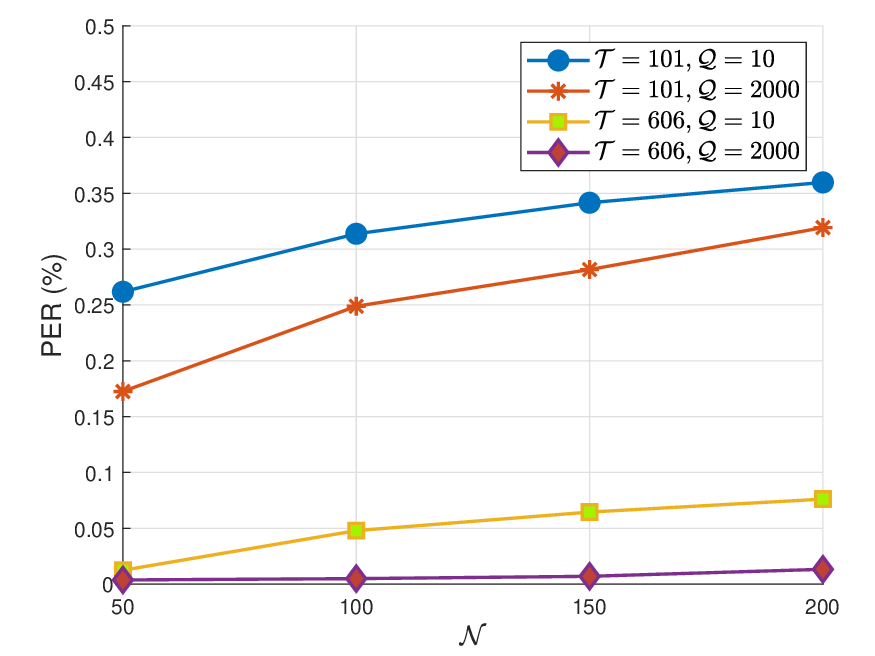}
    \caption{PER as a function of $\mathcal{N}$ with $\mathcal{D} = 2$ km and different values of $\mathcal{Q}$, $\mathcal{T}$ and $\mathcal{R}=2$.}
    \label{fig:per_rty_a}
\end{figure}

\begin{figure}[!t]
    \centering
    \includegraphics[width=1.0\columnwidth]{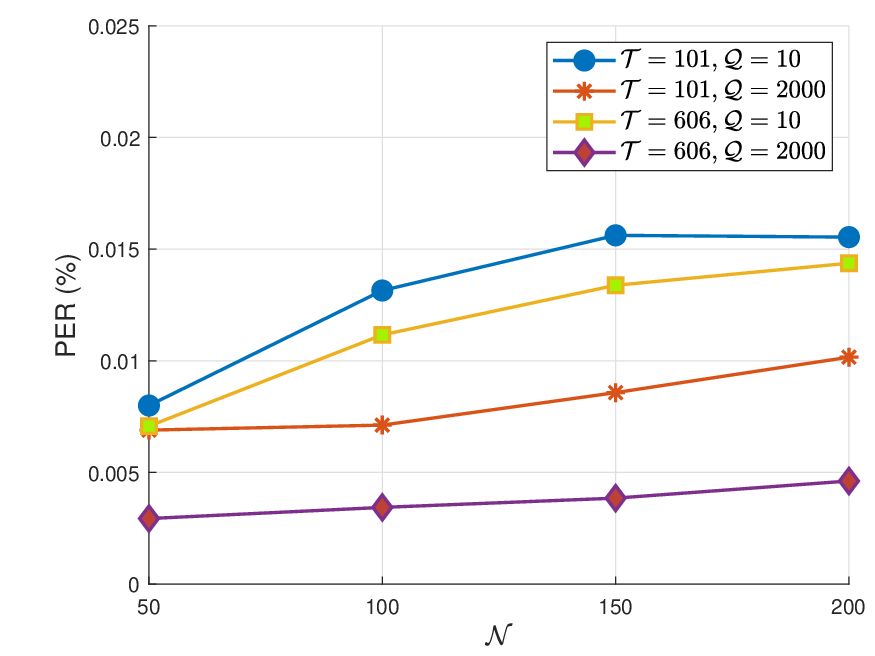}
    \caption{PER as a function of $\mathcal{N}$ with $\mathcal{D} = 2$ km and different values of $\mathcal{Q}$, $\mathcal{T}$ and $\mathcal{R}=200$.}
    \label{fig:per_rty_b}
\end{figure}

Figs.~\ref{fig:per_rty_a} and \ref{fig:per_rty_b} depict the PER as a function of the number of nodes ($\mathcal{N}$) for various combinations of slotframe length ($\mathcal{T}$) and queue size ($\mathcal{Q}$), while considering different values for the maximum number of retries, denoted as $\mathcal{R}$. As we observe, increasing both $\mathcal{N}$ and $\mathcal{T}$ results in a reduction in PER, leading to improved performance in both figures. Notably, from the results shown in Fig.~\ref{fig:per_rty_a} (with $\mathcal{R} = 2$), it is evident that increasing the slotframe length has a more pronounced impact on reducing PER compared to increasing the transmission queue. In addition, increasing the queue size $200$ times, from $\mathcal{Q} = 10$ to $\mathcal{Q} = 2000$, resulted in a reduction of up to $1.52$ times for $\mathcal{T}=101$ and up to $12$ times for $\mathcal{T}=606$. Meanwhile, a sixfold increase in the slotframe length, from $101$ to $606$, led to a PER reduction of up to $21$ times for $\mathcal{Q}=10$ and up to $55$ times for $\mathcal{Q}=2000$. 

Fig.~\ref{fig:per_rty_b} reveals that, with $\mathcal{R} = 200$, the queue size has a more pronounced impact on PER reduction than the slotframe length. Specifically, increasing the queue size $200$ times, from $\mathcal{Q} = 10$ to $\mathcal{Q} = 2000$, reduced the PER by up to $1.85$ times for $\mathcal{T}=101$ and up to $3.47$ times for $\mathcal{T}=606$. Furthermore, a six-unit increase in the length of the slotframe, from $101$ to $606$, resulted in a reduction of up to $1.18$ times for $\mathcal{Q}=10$ and up to $2.34$ times for $\mathcal{Q}=2000$ in PER. Finally, comparing Figs.~\ref{fig:per_rty_a} and \ref{fig:per_rty_b}, we observe that increasing the number of retransmissions yields a more significant PER decrease.

\begin{figure}[!t]
    \centering
    \includegraphics[width=1.0\columnwidth]{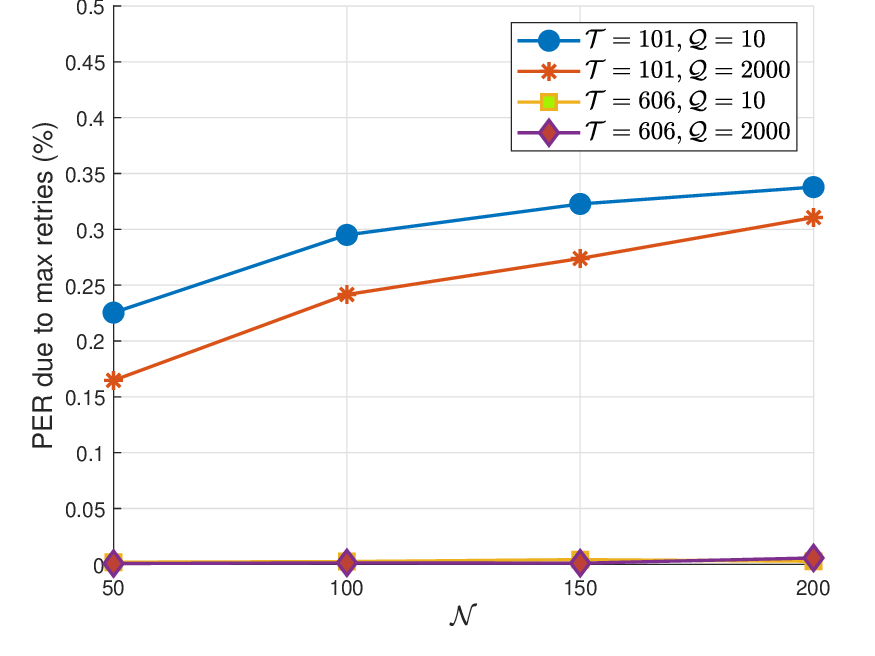}
    \caption{PER due to maximum retries as a function of $\mathcal{N}$ with $\mathcal{D} = 2$ km and $\mathcal{R} = 2$.}
    \label{fig:perc_maxret_perda_2}
\end{figure}

In the sequel, Fig.~\ref{fig:perc_maxret_perda_2} analyzes the PER as a function of $\mathcal{N}$, with special focus on errors caused by the maximum number of retries being reached considering $\mathcal{R} = 2$. We notice that a slotframe length of $\mathcal{T} = 101$ exhibits a higher PER for the maximum number of retries. This is because, for $\mathcal{T}=606$, there is a lower number of collisions, decreasing both the need for retransmission and the number of packets failing due to the retries limit. Additionally, it is noteworthy that this result indicates the potential occurrence of more losses related to transmission queues. Regarding the increase in the message queue from $\mathcal{Q} = 10$ to $\mathcal{Q} = 2000$, simulation results demonstrate a lower PER. This confirms that increasing the message queue influences transmission success in scenarios with an extended slotframe length. 

\begin{figure}[!t]
    \centering
    \includegraphics[width=1.0\columnwidth]{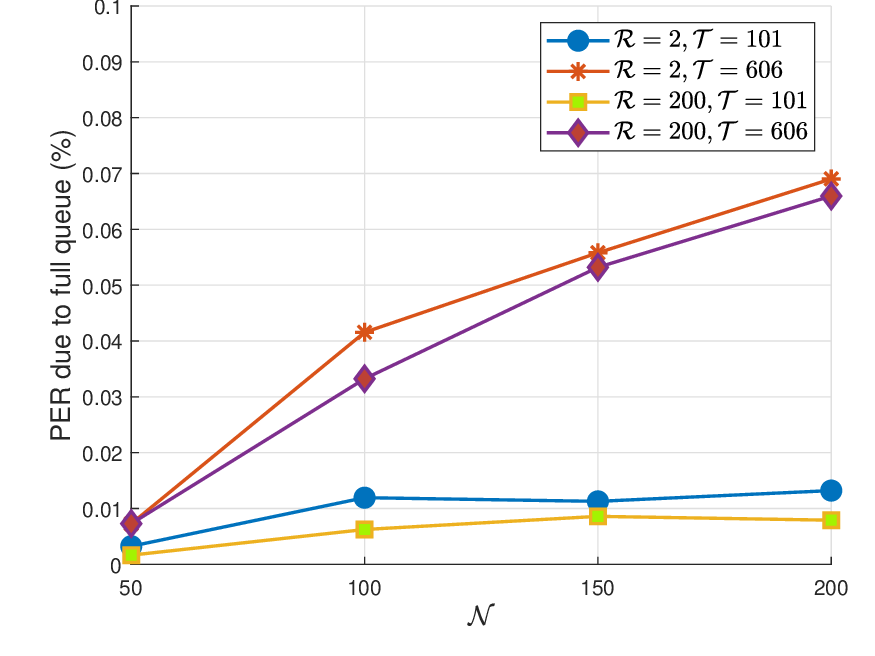}
    \caption{PER due to full queue as a function of $\mathcal{N}$ with $\mathcal{D} = 2$ km and $\mathcal{Q} = 10$.}
    \label{fig:perc_quef_perda_2}
\end{figure}

Next, Fig.~\ref{fig:perc_quef_perda_2} analyzes the PER as a function of $\mathcal{N}$, with a specific focus on errors caused due to the full queue with $\mathcal{Q} = 10$. Notably, a lower full queue error rate is achieved with reduced $\mathcal{T}$ values. This behavior can be attributed to the fact that, with $\mathcal{T} = 101$, the frequency of errors due to reaching the maximum number of retries is higher than with $\mathcal{T} = 606$. However, as previously observed in Fig. \ref{fig:perc_maxret_perda_2}, the increase in $\mathcal{T}$ contributes to a more efficient allocation and distribution of slots compared to $\mathcal{T} = 101$. Consequently, timely message transmission is facilitated by reducing errors due to the maximum number of retransmissions at $\mathcal{T} = 606$, albeit resulting in an accumulation of errors due to the full message queue. In a scenario with $\mathcal{T} = 606$, an increase in $\mathcal{R}$, as expected, reduces the error rate due to a full queue for both $\mathcal{T} = 101$ and $\mathcal{T} = 606$. Specifically, simulations with $\mathcal{R} = 200$ and $\mathcal{T} = 606$ exhibit a decrease of up to $19\%$ in the full queue error rate compared to $\mathcal{R} = 2$ and $\mathcal{T} = 606$, which can be attributed to the additional number of retransmission attempts. Similarly, for simulations with $\mathcal{R} = 200$ and $\mathcal{T} = 101$, a reduction of up to $50\%$ in the full queue error rate is observed compared to $\mathcal{R} = 2$ and $\mathcal{T} = 101$, further justified by the higher number of transmission attempts.

\begin{figure}[!t]
    \centering
    \includegraphics[width=1.0\columnwidth]{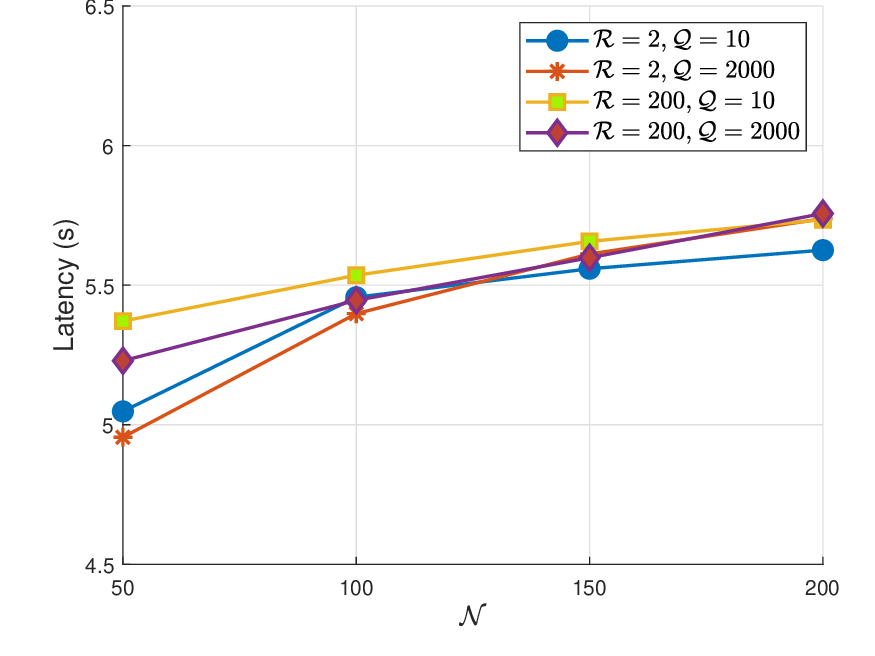}
    \caption{Latency as a function of $\mathcal{N}$ for different values of $\mathcal{R}$ and $\mathcal{Q}$ with $\mathcal{D} = 2$ km and $\mathcal{T} = 101$.}
    \label{fig:latencia_101}
\end{figure}

\begin{figure}[!t]
    \centering
    \includegraphics[width=1.0\columnwidth]{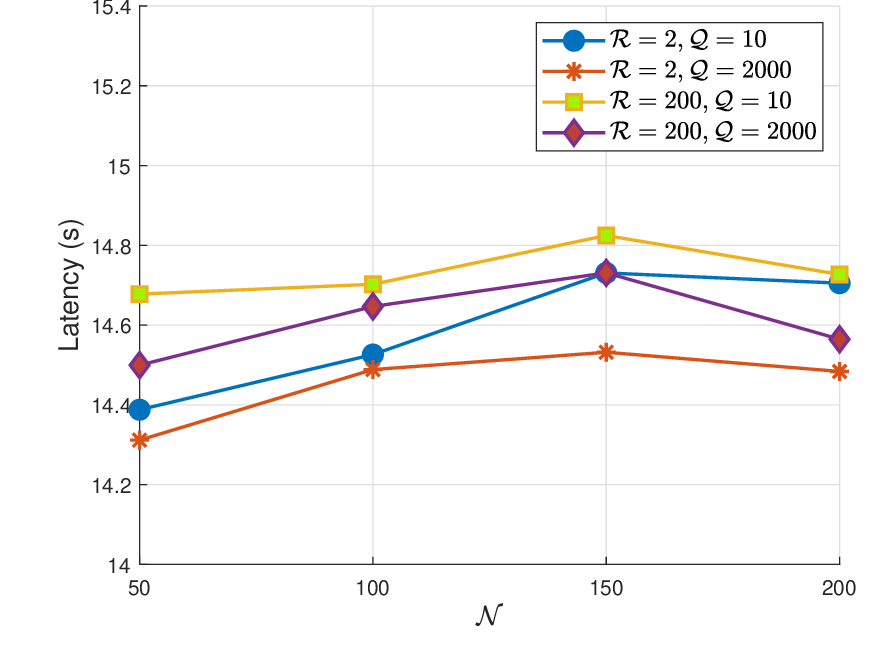}
    \caption{Latency as a function of $\mathcal{N}$ for different values of $\mathcal{R}$ and $\mathcal{Q}$ with $\mathcal{D} = 2$ km and $\mathcal{T} = 606$.}
    \label{fig:latencia_606}
\end{figure}

Finally, the results depicted in Fig.~\ref{fig:latencia_101} illustrate the latency variation versus $\mathcal{N}$ when $\mathcal{T} = 101$. The figure reveals that an increase in both $\mathcal{N}$ and $\mathcal{R}$ leads to a slight elevation in latency. This increase can be attributed to the need for retransmitting packets across multiple hops in the 6TiSCH mesh topology. Specifically, at low values of $\mathcal{N}$, an enlarged queue size prevents unnecessary packet retransmissions while facilitating efficient slot allocation. However, as $\mathcal{N}$ increases, this efficient allocation diminishes, resulting in higher-latency packet delivery. It is worth mentioning that, nonetheless, the impact of a slightly increased latency depends on the application of the network. For instance, for systems with milder latency requirements, the increase in $\mathcal{Q}$ is usually beneficial since it reduces the PER. Furthermore, the variation in latency as a function of $\mathcal{N}$ when $\mathcal{T} = 606$ can be observed in Fig.~\ref{fig:latencia_606}. As depicted in Fig.~\ref{fig:latencia_101}, increasing $\mathcal{R}$ results in higher latency in Fig.~\ref{fig:latencia_606}, primarily due to the multiple-hop retransmissions inherent in the mesh topology. However, contrary to Fig.~\ref{fig:latencia_101}, latency decreases with increasing $\mathcal{Q}$ for all values of $\mathcal{N}$. This behavior arises because a larger slotframe can more effectively utilize a higher $\mathcal{Q}$, ultimately resulting in lower latency. For $\mathcal{T} = 101$ the maximum tree depth is $4$ and $6$ for, respectively, $\mathcal{N} = 50$ and $\mathcal{N} = 200$, and for $\mathcal{T} = 606$ the maximum tree depth is $6$ and $8$ for, respectively, $\mathcal{N} = 50$ and $\mathcal{N} = 200$, which  justifies the increasing latency with $\mathcal{T}$. 

\section{Conclusions}
\label{sec4}
In this work, we investigated the influence of 6TiSCH parameters by varying three key factors: the maximum number of retries, the transmission queue size, and slotframe length. The results demonstrate that increasing the transmission queue size and the maximum number of retries leads to a reduction in terms of PER. However, it's worth noting that the impact of increasing the maximum retries is significantly more pronounced than that of increasing the transmission queue size. Furthermore, when the slotframe length is greater, increasing the transmission queue size is more effective in achieving both latency reduction and PER improvement. In contrast, for smaller slotframe, although both latency and PER decrease with fewer nodes, there is an increase in latency as the number of nodes grows.

\normalsize

\bibliographystyle{IEEEtran}
\bibliography{biblio}

\begin{thebibliography}{10}
\providecommand{\url}[1]{#1}
\csname url@samestyle\endcsname
\providecommand{\newblock}{\relax}
\providecommand{\bibinfo}[2]{#2}
\providecommand{\BIBentrySTDinterwordspacing}{\spaceskip=0pt\relax}
\providecommand{\BIBentryALTinterwordstretchfactor}{4}
\providecommand{\BIBentryALTinterwordspacing}{\spaceskip=\fontdimen2\font plus
\BIBentryALTinterwordstretchfactor\fontdimen3\font minus
  \fontdimen4\font\relax}
\providecommand{\BIBforeignlanguage}[2]{{%
\expandafter\ifx\csname l@#1\endcsname\relax
\typeout{** WARNING: IEEEtran.bst: No hyphenation pattern has been}%
\typeout{** loaded for the language `#1'. Using the pattern for}%
\typeout{** the default language instead.}%
\else
\language=\csname l@#1\endcsname
\fi
#2}}
\providecommand{\BIBdecl}{\relax}
\BIBdecl

\bibitem{Chettri.IoT.2020}
L.~Chettri and R.~Bera, ``A comprehensive survey on {I}nternet of {T}hings
  ({IoT}) toward {5G} wireless systems,'' \emph{IEEE Internet Things J.},
  vol.~7, no.~1, pp. 16--32, 2020, doi: {10.1109/JIOT.2019.2948888}.

\bibitem{Keersmaeker.2023}
F.~De~Keersmaeker, Y.~Cao, G.~K. Ndonda, and R.~Sadre, ``A survey of public
  {IoT} datasets for network security research,'' \emph{IEEE Commun. Surv.
  Tutor.}, vol.~25, no.~3, pp. 1808--1840, 2023, doi:
  \url{10.1109/COMST.2023.3288942}.

\bibitem{Wang.2022}
Z.~Wang, D.~Liu, Y.~Sun, X.~Pang, P.~Sun, F.~Lin, J.~C.~S. Lui, and K.~Ren, ``A
  survey on {IoT}-enabled home automation systems: Attacks and defenses,''
  \emph{IEEE Commun. Surv. Tutor.}, vol.~24, no.~4, pp. 2292--2328, 2022, doi:
  \url{10.1109/COMST.2022.3201557}.

\bibitem{Venturini.2022}
L.~F. Venturini, Y.~L. Baracy, R.~V. P.~C. Silva, B.~K. De~Freitas, E.~P.
  Dos~Santos, N.~O. Branco, M.~E.~P. Monteiro, J.~F. Hübner, and D.~Issicaba,
  ``Balancing decentralization for restoration in power distribution systems
  with agents,'' \emph{IEEE Access}, vol.~10, pp. 77\,993--78\,001, 2022, doi:
  \url{10.1109/ACCESS.2022.3192847}.

\bibitem{Yadav.2023}
R.~Yadav, R.~Saroj, A.~Kumar~Verma, and A.~Kumar~Mishra, ``A survey of {IoT}
  and machine learning based monitoring of the growth of crops using blockchain
  technology,'' in \emph{Int. Conf. on IoT, Comm. and Autom. Tech. (ICICAT)},
  2023, pp. 1--7, doi: \url{10.1109/ICICAT57735.2023.10263755}.

\bibitem{Melo.2023}
L.~S. Melo, F.~L. Tofoli, D.~Issicaba, M.~E.~P. Monteiro, G.~C. Barroso, R.~F.
  Sampaio, and R.~P.~S. Leao, ``Co-simulation platform for the assessment of
  transactive energy systems,'' \emph{Electr. Pow. Syst. Res.}, vol. 223, p.
  109693, 2023, doi: \url{https://doi.org/10.1016/j.epsr.2023.109693}.

\bibitem{Kim.2017}
K.~T. {Kim}, H.~{Kim}, H.~{Park}, and S.~{Kim}, ``An industrial {IoT MAC}
  protocol based on {IEEE 802.15.4e TSCH} for a large-scale network,'' in
  \emph{Int. Conf. on Advanced Commun. Technol. (ICACT)}, 2017, pp. 721--724,
  doi: \url{10.23919/ICACT.2017.7890187}.

\bibitem{Barros.2022}
J.~L.~V. {de Barros}, M.~E.~P. Monteiro, G.~S. {Peron}, G.~L. Moritz, O.~K.
  Rayel, and R.~D. Souza, ``{LoRaWAN} vs. {6TiSCH}: Which one scales better?''
  \emph{Comput. Commun.}, vol. 184, pp. 1--11, 2022, doi:
  \url{https://doi.org/10.1016/j.comcom.2021.12.004}.

\bibitem{Accettura.2015}
N.~{Accettura}, E.~{Vogli}, M.~R. {Palattella}, L.~A. {Grieco}, G.~{Boggia},
  and M.~{Dohler}, ``Decentralized traffic aware scheduling in {6TiSCH}
  networks: Design and experimental evaluation,'' \emph{IEEE Internet Things
  J.}, vol.~2, no.~6, pp. 455--470, 2015, doi: \url{10.1109/JIOT.2015.2476915}.

\bibitem{Watteyne.2015}
T.~{Watteyne}, C.~{Adjih}, and X.~{Vilajosana}, ``Lessons learned from
  large-scale dense {IEEE802.15.4} connectivity traces,'' in \emph{IEEE Int.
  Conf. on Automation Science and Engineering (CASE)}, 2015, pp. 145--150, doi:
  \url{10.1109/CoASE.2015.7294053}.

\bibitem{Municio.2019}
E.~Municio, G.~Daneels, M.~Vucinic, S.~Latré, J.~Famaey, Y.~Tanaka, K.~Brun,
  K.~Muraoka, X.~Vilajosana, and T.~Watteyne, ``Simulating {6TiSCH} networks,''
  \emph{Trans. Emerg. Telecommun. Technol.}, vol.~30, no.~3, p. e3494, 2019,
  doi: \url{10.1002/ett.3494}.

\bibitem{IEEE.2012}
``{IEEE} standard for local and metropolitan area networks--part 15.4: Low-rate
  wireless personal area networks (lr-wpans) amendment 3: Physical layer (phy)
  specifications for low-data-rate, wireless, smart metering utility
  networks,'' \emph{{IEEE} Std 802.15.4g-2012}, pp. 1--252, 2012, doi:
  \url{10.1109/JPROC.2010.2070470}.

\bibitem{Goldsmith.05}
A.~Goldsmith, \emph{Wireless Communications}.\hskip 1em plus 0.5em minus
  0.4em\relax Cambridge University Press, 2005, isbn: \url{978-0521837163}.

\bibitem{CC1352.2020}
``{CC1352R} simplelink high-performance multi-band wireless {MCU},'' 2020, url:
  \url{https://www.ti.com/lit/gpn/cc1352p}.

\bibitem{Elsts.2020}
A.~Elsts, ``{TSCH-Sim}: Scaling up simulations of {TSCH} and {6TiSCH}
  networks,'' \emph{Sensors}, vol.~20, no.~19, 2020, doi:
  \url{10.3390/s20195663}.

\bibitem{RFC8180}
X.~Vilajosana, K.~Pister, and T.~Watteyne, ``{Minimal {IPv6} over the {TSCH}
  Mode of {IEEE} 802.15.4e ({6TiSCH}) Configuration},'' Internet Requests for
  Comments, {RFC} 8180, May 2017, doi: \url{https://doi.org/10.17487/RFC8180}.

\bibitem{Palattella.2016}
M.~R. Palattella, T.~Watteyne, Q.~Wang, K.~Muraoka, N.~Accettura, D.~Dujovne,
  L.~A. Grieco, and T.~Engel, ``On-the-fly bandwidth reservation for {6TiSCH}
  wireless industrial networks,'' \emph{IEEE Sens. J.}, vol.~16, no.~2, pp.
  550--560, 2016, doi: \url{10.1109/JSEN.2015.2480886}.

\end{thebibliography}

\end{document}